\documentstyle[12pt]{article}

\author{M. Temple-Raston
\thanks{current address: Centre for Industrial Control, Mechanical
Engineering, Concordia University, Montreal, Quebec, Canada}\\
Department of Mathematics and Statistics,\\
Concordia University,\\
Montr\'eal, Qu\'ebec, Canada}
\title{Solitonic photons and intermediate vector bosons
}

\begin{document}

\maketitle
\begin{abstract}
A four-dimensional topological field theory is introduced which
generalises $B\wedge F$ theory to give a Bogomol'nyi structure.
A class of non-singular, finite-Action, stable solutions to the
variational field equations is identified.  The solitonic solutions
are analogous to the instanton in Yang-Mills theory. The solutions
to the Bogomol'nyi equations in the topologically least complicated
$U(1)$ theory have a covariant phase space of dimension four---the
same as that for photons. The dimensional reduction of the
four-dimensional Lagrangian is also examined. Bogomol'nyi $U(2)$
solitons resembling the intermediate vector bosons
$Z_o$, $W^{\pm }$ are identified.
\end{abstract}

\section{Introduction}

The photon is a non-singular, finite-energy field configuration with a
definite frequency $\nu $, wave vector ${\bf k}$, and helicity $\pm 1$.
Furthermore, the energy of the photon, $E$, and its momentum, ${\bf p}$,
satisfy the dispersion relation $E=c\mid {\bf p}\mid $. The dispersion
relation implies that the phase space, ${\cal P}$, exists and must
presumably be a well-behaved manifold (e.g., Hausdorff). If the photon
is solitonic (in the general sense of the word), then the phase space
becomes the `covariant' phase space. The covariant phase space is given
by the moduli space of solutions to the photonic time-dependent field
equations. At present, field equations for a solitonic photon are
unknown. However, it is a strong requirement on the possible field
equations to have a well-behaved covariant phase space, ${\cal P}$.
If additionally we require that the photonic covariant phase space be
finite dimensional and in particular of dimension four, we now possess
very strong conditions on classical field theories which dictate the
terms under which a theory is acceptable. Subject to these physical
requirements on the covariant phase space, we introduce in this paper
a Lagrangian field theory which possesses Bogomol'nyi solitons with
properties similar to the photon. The theory also contains stationary
solitons that resemble other force mediating particles. They occur as
Bogomol'nyi solitons to the time-reduced field equations.

We now summarize the sections of this paper. In the next section we
introduce two classical, generally covariant Lagrangian field theories
defined on a four-manifold. The natural Bogomol'nyi structures are
identified. Non-singular solutions to the Bogomol'nyi equations are shown
to be stable under small perturbations. When the theory is defined on a
compact, complex K\"ahler surface a finite dimensional covariant phase
space is found for the Bogomol'nyi solitons, with good topological
behaviour.  The covariant phase space for abelian (diagonal)
Bogomol'nyi solitons over various compact, complex surfaces is examined
in detail. In section three we dimensionally reduce the four-dimensional
topological field theory using a time-symmetry to a three-dimensional
topological field theory. The three-dimensional field theory permits
electric, magnetic and dyonic monopoles. From the general arguments
in \cite{olive} the particle spectrum and classical mass compare
favourably with that of the intermediate vector bosons.

\section{Topological solitons}

Let $\pi :P\rightarrow M$ be a principal $G$-bundle over an oriented,
compact, connected four manifold $M$, and denote by $E=P\times _G{\cal G}$,
the associated adjoint vector bundle where ${\cal G}$ is the Lie algebra
of $G$. Denote by ${\cal A}(P)$ the space of connections on $P$.
Let $A,B\in {\cal A}(P)$. We introduce local coordinate indices
$\mu =0,\dots ,3$ on $M$. The ${\cal G}$-valued connections or vector
potentials, $A_\mu $ and $B_\mu $, induce exterior covariant derivatives
$D_\mu ^A=\partial _\mu+A_\mu $ and $D_\mu ^B=\partial _\mu +B_\mu $
on the associated adjoint vector bundle $E$. The curvatures $H$ and $K$
are defined by $2D_{[\mu}^AD_{\nu ]}^As=sH_{\mu \nu }$ and
$2D_{[\mu }^BD_{\nu ]}^Bs=sK_{\mu \nu }$.
As a result, $H$ and $K$ are two-forms on $M$ taking values in $E$. We
introduce a Lagrangian Action functional constructed in such a way to
produce invariants and to be geometrically meaningful. Physically, the
invariants and geometrical interpretation allow us to argue the existence
and stability of our solitonic solutions. The Lagrangian theory is given
by the functional
\begin{equation}
\label{Lang}
\begin{array}{c}
{\cal L}_{\pm }(A,B)=\int_MH_{[\mu \nu }^aK_{\lambda \rho ]}^b\
{\rm tr}(T^a){\rm tr}(T^b)\ d^4x \\ \pm \frac 12\int_MK_{[\mu \nu }^a
K_{\lambda \rho]}^b\ {\rm tr}(T^aT^b)\ d^4x.
\end{array}
\end{equation}
The generators of the Lie algebra are denoted by $T^a$. For gauge groups
where ${\rm tr}(T^a)=0$, the Lagrangian reduces to the second
integral---these are the topological field theories studied by Baulieu
and Singer \cite{singer}. We do not a priori restrict the Lie group.

The variational field equations are
\begin{equation}
\label{field}D_{[\mu }^AK_{\nu \rho ]}=0,\qquad
D_{[\mu }^BH_{\nu \rho ]}=0.
\end{equation}
The Lagrangian can be rewritten as
\begin{equation}
\label{Lanbog}
\begin{array}{c}
2
{\cal L}_{\pm }=\pm \int_M<(H_{[\mu \nu }\otimes I_E\pm I_E\otimes
K_{[\mu\nu })(H_{\lambda \rho ]}\otimes I_E\pm I_E\otimes
K_{\lambda \rho ]})\ >\ d^4x \\ \mp \int_MH_{[\mu \nu }^a
H_{\lambda \rho ]}^b\ {\rm tr}(T^aT^b)\
d^4x.
\end{array}
\end{equation}
$I_E$ denotes the identity transformation on the adjoint bundle, $E$. The
Lagrangian in (\ref{Lanbog}) is now in Bogomol'nyi form. We need to
interpret the first term in ${\cal L}_{\pm }$. Let $E_A$ and $E_B$ be the
adjoint vector bundles equipped with $D_\mu ^A$ or $D_\mu ^B$, respectively.
$E_B^{*}$ is the dual bundle to $E_B$. Now, recall that the curvature of the
tensor product bundle $E_A\otimes E_B^{*}$ is given by $(\Omega _{E_A\otimes
E_B^{*}})_{\mu \nu }=$$H_{\mu \nu }\otimes I_E-I_E\otimes K_{\mu \nu }$ \cite
{kob}. Likewise, the curvature of $E_A\otimes E_B$ is given by $(\Omega
_{E_A\otimes E_B})_{\mu \nu }=$$H_{\mu \nu }\otimes I_E+I_E\otimes K_{\mu
\nu }$. Since the trace inner product on $E$ generalizes naturally to an
inner product on $E\otimes E$ and $E\otimes E^{*}$, we have signified both
inner products by $<\ >$ in (\ref{Lanbog}). The appropriate inner product is
determined by the context. Under an infinitesimal perturbation of the vector
potential, the integrand in the first term of both Lagrangians
(\ref{Lanbog}) changes by an exact form. The exact form vanishes when
integrated over compact $M$, so that the first term in (\ref{Lanbog}) is
invariant under small perturbations.

In our view, the more interesting theory is that of the Lagrangian
${\cal L}_{-}$, so we concentrate on it. The Bogomol'nyi equations in
(\ref{Lanbog}) for ${\cal L}_{-}$ are seen to be a vanishing curvature
condition on the tensor product bundle $E_A\otimes E_B^{*}$:
\begin{equation}
\label{bog4}(\Omega _{E_A\otimes E_B^{*}})_{\mu \nu }=H_{\mu \nu }\otimes
I_E-I_E\otimes K_{\mu \nu }=0.
\end{equation}
Solutions to (\ref{bog4}) automatically satisfy the variational field
equations (\ref{field}). The equations (\ref{bog4}) give useful information
about $H$ and $K$; they imply that $H$ and $K$ are projectively flat,
\begin{equation}
\label{bog4'}H_{\mu \nu }=K_{\mu \nu }=iF_{\mu \nu }I,
\end{equation}
where $F$ is a real-valued two form on $M$. The projective flatness reduces
the gauge group down to $U(n)$. Non-trivial solitonic solutions $(A_\mu
,B_\mu )$ to the Bogomol'nyi equations (\ref{bog4}) are stable under a small
perturbation when the Lagrangian Action is invariant and
\begin{equation}
\label{stable}\int_M{\rm tr}(H_{[\mu \nu }H_{\lambda \rho ]})\ d^4x
\end{equation}
is non-vanishing.

There are clear similarities between what we have constructed above and the
(anti-)self-dual instanton in Yang-Mills theory. We, however, will not call
these solutions `instantons'. Because physicists have come to associate
instantons with Euclidean four-space---only trivial instanton solutions
exist in Minkowski space. Unlike instantons, the Bogomol'nyi solutions in
these topological field theories are not affected by the metric signature of
space-time. In light of this, we call non-trivial, stable, non-singular
solutions to the Bogomol'nyi equations (\ref{bog4'}) topological solitons.
The term `soliton' is used rather generally in this paper, and does not
suggest that the solutions are connected to inverse scattering or that
soliton-soliton interactions are trivial. Note that a space-time metric is
not needed to define the Lagrangian field theory (\ref{Lang}). Therefore the
underlying manifold $M$ need not be space-time at all. In view of this, we
define $M$ to be a four-dimensional manifold which defines the {\it topology}
of the field configuration on space-time. The topology is defined by the
number of isolated zeros and the multiplicities in the field configuration
\cite{chern}. The topology of $M$, call it the topological field manifold,
defines a set of topologically equivalent field configurations on
space-time. If $M$ is indeed endowed with the space-time metric and viewed
as space-time, then this reduces to a particular example of the general
theory we propose.

We now restrict to those topological solitons $(A_\mu ,B_\mu )$ of the
Bogomol'nyi equations (\ref{bog4}) for which $A_\mu =B_\mu $. Let us call
these solutions `diagonal topological solitons'. The variational field
equations (\ref{field}) become the Bianchi identities. The only other
condition on the vector potential, $A_\mu $, comes from the Bogomol'nyi
equations, which require that $A_\mu $ be projectively flat. We must now
state our connection to Maxwell's theory of electromagnetism. Consider first
a $U(1)$ gauge theory. Since in a $U(1)$ gauge theory projective flatness is
a vacuous requirement, the abelian vector potential is completely
unconstrained. We are led therefore to introduce Maxwell's equations. We do
this by requiring that $F_{\mu \nu }$ in equation (\ref{bog4'}) is a Maxwell
field (Faraday tensor). The Bianchi identities for $H$ and $K$ imply
automatically that $dF=0$. The second half of the Maxwell equations,
\thinspace $^{*}d^{*}F=j$, however, we must introduce by hand. The
introduction of Maxwell's equations at the level of $F_{\mu \nu }$ implies
that in $U(n)$, $n>1$, gauge theories $H_{\mu \nu }=iF_{\mu \nu }I$
satisfies the non-homogeneous Yang-Mills equations, $^{*}D^A\ ^{*}H=J\equiv
jI$. In the following, we permit the second half of the Maxwell equations be
relaxed if needed.

So far, the Lagrangian introduced above has yielded a Bogomol'nyi structure.
The Bogomol'nyi structure suggests that interesting particle-like solutions
to the theory might be found. However, we have also seen that in the case of
the abelian gauge group only vacuous conditions remain in the theory.
Nothing further remains in our theory to determine these solitons. The way
forward is made clearer by the non-abelian theory. The Bogomol'nyi equations
(\ref{bog4'}) are non-trivial in the non-abelian theory; they require that
the vector potential be projectively flat. Projectively flat vector
potentials are closely related to Einstein-Hermitian connections, which in
turn express in differential geometry the algebraic geometric concept of
`stable vector bundle' \cite{kob}. As a result of this close connection to
stable vector bundle, projectively flat vector potentials often have a
well-behaved moduli space---like self-dual instantons and BPS magnetic
monopoles. In the four-dimensional theory studied in this section, the
moduli space of topological solitons is equivalent to the covariant phase
space (or, the space of motions).

To establish the link between projectively flat and Einstein-Hermitian
connections, we must introduce further geometrical structure into the
theory. We require that the topological field manifold $M$ be a compact,
complex K\"ahler surface. Clearly we have introduced the K\"ahler structure
$(g,\Phi )$ on $M$ at the expense of the space-time structure. $M$ is no
longer space-time. The K\"ahler metric and K\"ahler form are given by $g$
and $\Phi $, respectively. We also restrict the $U(1)$ theory, which is all
that interests us in this section, to holomorphic Hermitian line bundles
$(L,\bar \partial ,h=\ <\ >)$ over $M$. The set of connections compatible
with the added geometric structure (the K\"ahler metric $g$, the complex
structure $\bar \partial $) are denoted by ${\cal D}^{1,1}(L,h)$.
Einstein-Hermitian connections are defined as the connections with constant
mean curvature, $K(A)=c$, compatible with the holomorphic Hermitian
structure. Let ${\cal E}(L,h)$ denote the set of all Einstein-Hermitian
connections in ${\cal D}^{1,1}(L,h)$. Kobayashi has shown that all
connections in ${\cal D}^{1,1}(L,h)$ are Einstein-Hermitian up to a
conformal transformation of the Hermitian structure, $h$\cite{kob1}. The
constant $c$ is given by
$$
c=\frac{2\pi }V\deg (L),
$$
where $V={\rm Vol}(M)$ and $\deg (L)=\int_Mc_1(L)\wedge \Phi $. $c$ is a
topological invariant when $M$ is K\"ahler (when $\Phi $ is closed), and
depends only on the cohomology classes of $\Phi $ and $c_1(L)$ \cite{kob}.

As mentioned before, the moduli space of Einstein-Hermitian connections is
equivalent to the covariant phase space for the topological solitons defined
by the Bogomol'nyi equations (\ref{bog4'}) on $(M,g,\Phi )$. The covariant
phase space has been studied extensively by H.J. Kim \cite{kim}. Kim proves
that the covariant phase space ${\cal M}=E(L,h)/U(1)$ is a non singular
K\"ahler manifold if $H^2(M,{\rm End}^0(L))=0$, and
\begin{equation}
\label{degM}\deg (M)\equiv \int_Mc_1(M)\wedge \Phi \geq 0.
\end{equation}
When the inequality (\ref{degM}) is satisfied, the complex dimension of the
covariant phase space of abelian diagonal topological solitons is found to
be
\begin{equation}
\label{dimen}\dim {}_{{\bf C}}(M)=q,
\end{equation}
where $q\equiv h^{(0,1)}$ is the irregularity of $M$. The irregularity is
one of the Hodge numbers in complex cohomology theory, defined generally
by $h^{(p,q)}\equiv \dim {}_{{\bf C}}H^{p,q}(M)$. The natural symplectic
structure on the covariant phase space is inherited from the K\"ahler
structure on the configuration manifold, $M$:
$$
\Theta (a,b)=\int_M{\rm tr}(a\wedge b)\wedge \Phi ,
$$
where $a,b\in T{\cal M}$ \cite{mukai}.

Which field configuration topologies lead to interesting covariant phase
spaces in this theory? The simplest topology, that of the K3 surface, has
irregularity $q=0$. The dimension of the covariant phase space with this
field topology is therefore zero, from equation (\ref{dimen}). This phase
space is too small to be interesting. The next simplest field
configuration, that of the flat complex two-torus, $T^2={\bf C}^2/\Gamma $,
has irregularity $q=2$. For this field configuration the real dimension of
the covariant phase space is four. While the dimension of the covariant
phase space is non-zero, the stability of the topological solitons under a
small perturbation is still under question. From (\ref{stable}), we recall
that the stability for vector bundles of rank $r>1$ is assured when the
second Chern class is non-zero. For line bundles, however, equation
(\ref{stable}) is not generally a topological invariant. We can make
(\ref{stable}) a topological invariant by interpreting $F_{\mu \nu }$ in
the Bogomol'nyi equation (\ref{bog'}) as a curvature on the tangent bundle,
$TM$. Equation (\ref{stable}) is then proportional to the signature of the
underlying four-manifold, $M$. Thus, the stability of the topological
soliton in the abelian theory is argued strictly from the topology of
the topological field manifold, $M$. Returning to the flat complex
two-torus, recall that for $T^2$ the irregularity ($q$) and the geometric
genus ($p_g$) take the values $q=2$ and $p_g=1$. Also $c_1^2(T^2)=0$.
{}From the Riemann-Roch formula
$$
c_1^2(M)+c_2(M)=12(1-q+p_g)
$$
we conclude that the signature vanishes for $T^2$. Therefore solutions to
the Bogomol'nyi equations with a topological field manifold $M=T^2$ are
unstable. The next topological field manifold we examine is found in the
Enriques-Kodaira classification of compact, complex surfaces (free from
exceptional curves). There are two possibilities: minimal elliptic surfaces
and minimal algebraic surfaces of `general type'. We examine here the
simpler of the two. Let $M$ be a minimal elliptic ($c_1^2(M)=0$) surface
with irregularity $q=2$. Again, the covariant phase space is of real
dimension four. But in addition now, stability is assured for any
topological field manifold $M$ with geometric genus $p_g>1$. Thus we have
shown that the real dimension of the covariant phase space for $U(1)$
diagonal topological solitons (understood to be stable) on minimal elliptic
complex surfaces with geometric genus greater than one, is four. Recall
that the photon has a phase space (space of motions) of real dimension
four as well.

The explicit construction of topological solitons on minimal elliptic
topological field manifolds, $M$, is probably a difficult problem---as is
the case here, it is often much easier to construct the moduli space. A
connection between the (diagonal) topological soliton and the photon is
further suggested by a natural mechanism by which this theory has no
anti-solitons, in agreement with the notable absence of the anti-photon in
nature. This can be achieved if the transformations for time and parity
reversal of $H$ and $K$ are:
$$
\begin{array}{lrr}
T:H\longmapsto H & \qquad  & P:H\longmapsto -H \\
T:K\longmapsto -K & \qquad  & P:K\longmapsto -K
\end{array}
$$
Time reversal takes the Lagrangian Action ${\cal L}_{-}$ to
$-{\cal L}_{+}$ in equations (\ref{Lanbog}). The Bogomol'nyi equations
for $L_{+}$ are similar in appearance to those for $L_{-}$:
\begin{equation}
\label{bog+}(\Omega _{E_A\otimes E_B^{*}})_{\mu \nu }=H_{\mu \nu }\otimes
I_E+I_E\otimes K_{\mu \nu }=0.
\end{equation}
An index computation for these Bogomol'nyi equations yields the equations
$$
H_{\mu \nu }=-K_{\mu \nu }=iF_{\mu \nu }I.
$$
It is clear that diagonal topological solitons are no longer diagonal after
time reversal. On the other hand, under transformations of the parity the
Lagrangian Action is the same up to a overall sign---the mirror image of
the diagonal topological soliton is again a solution to the Bogomol'nyi
equations. This is compatible with two helicity states.

\section{Topological vector bosons}

In the previous section we studied the abelian topological field theory
defined by equation (\ref{Lang}). We found a Bogomol'nyi structure with
topological solitons that resemble photons. In this section we examine the
non-abelian theory. If indeed the topological solitons of section two can be
viewed as photons, then it is natural to expect other intermediate vector
bosons to be found in a similar manner. Our interest in this section now
lies in topological solitons with rest mass. Without loss of generality, we
can restrict our investigation to stationary topological solitons. At this
point one is reminded of the Montonen-Olive conjecture, which argues that
intermediate vector bosons should arise as solitons in a Lagrangian field
theory, analogous to the way that BPS magnetic monopole are found in
Yang-Mills-Higgs theory \cite{olive}. Since BPS magnetic monopoles and
instantons are related by a gauge symmetry, it is to be expected that we try
the same trick here on the topological solitons found in the previous
section. By assuming a gauge symmetry in time, the dimensionally reduced
Lagrangian Action gives the energy functional
${\cal E}(A,B,\Phi _A,\Phi _B)$ equal to
\begin{equation}
\label{TFT3}
\begin{array}{ccc}
\int_{M_3}K_{[\mu \nu }^a(D_{\rho ]}^B\Phi _B)^b\ {\rm tr}(T^aT^b) & - &
K_{[\mu \nu }^a(D_{\rho ]}^A\Phi _A)^b\
{\rm tr}(T^a){\rm tr}(T^b) \\ -H_{[\mu \nu }^a(D_{\rho ]}^B\Phi _B)^b\
{\rm tr}(T^a){\rm tr}(T^b) & + & (A\leftrightarrow B,\Phi _A\leftrightarrow
\Phi_B).
\end{array}
\end{equation}
We have followed the procedure of dimensional reduction in \cite{Forgacs}.
Note that in (\ref{TFT3}) there are two curvatures $H,K$ and two Higgs
fields $\Phi _A,\Phi _B$ corresponding to the two connections $A,B$. We
have added by hand the last term in (\ref{TFT3}) to symmetrise the energy
functional in the dependent fields. The symmetrisation although probably
desirable in principle in section two, made no change to the development of
diagonal topological solitons in the four-dimensional theory, so we left it
out. The energy functional looks rather complicated, but it does possess
analogous geometric structures to those found in the four dimensional
theory. The variational field equations arising from (\ref{TFT3}) are
\begin{equation}
\label{field3}
\begin{array}{c}
D^BH=0,\qquad \qquad D^BD^A\Phi _A=[H,\Phi _B], \\
D^AK=0,\qquad \qquad D^AD^B\Phi _B=[K,\Phi _A].
\end{array}
\end{equation}
The energy functional (\ref{TFT3}) can be rewritten as
\begin{equation}
\label{top3}
\begin{array}{ccc}
{\cal E}=\int_{M_3}<(H_{[\mu \nu }\otimes I_E-I_E\otimes K_{[\mu \nu }) &
\wedge & (D_{\rho ]}^A\Phi _A\otimes I_E-I_E\otimes D_{\rho ]}^B\Phi _B)>
\\ -\int_{M_3}<(D_{[\mu }^A\Phi _A\otimes I_E)\wedge (H_{\nu \rho ]}
\otimes I_E)> & + & (A\leftrightarrow B,\Phi _A\leftrightarrow \Phi _B).
\end{array}
\end{equation}
In this equation we begin to see the geometrical structure emerge out of
the energy functional (\ref{TFT3}). Let $E_A$ and $E_B$ be the vector
bundle $E$ equipped with either the connection $A$ or $B$, respectively.
Again, the curvature of the tensor product bundle $E_A\otimes E_B^{*}$ is
given by $\Omega _{E_A\otimes E_B^{*}}=$$H_{\mu \nu }\otimes I_E-I_E
\otimes K_{\mu \nu}$. This curvature expression appears in (\ref{top3}).
By defining $\Phi\equiv \Phi _A\otimes I_E-I_E\otimes \Phi _B$, then
$D_{E_A\otimes E_B^{*}}\Phi =D^A\Phi _A\otimes I_E-I_E\otimes D^B\Phi _B$,
and the first integral in (\ref{top3}) is therefore a topological
invariant. The Bogomol'nyi equations arising from equation (\ref{TFT3})
are
\begin{equation}
\label{bog}
\begin{array}{c}
H_{\mu \nu }\otimes I_E=I_E\otimes K_{\mu \nu } \\
D_\mu ^A\Phi _A\otimes I_E=I_E\otimes D_\mu ^B\Phi _B
\end{array}
\end{equation}
The first equation in (\ref{bog}) is a zero curvature condition on the
tensor product bundle $E_A\otimes E_B^{*}$. From an indices computation
we conclude that
\begin{equation}
\label{bog'}
\begin{array}{c}
H_{\mu \nu }=K_{\mu \nu }=iF_{\mu \nu }I_E, \\
D_\mu ^A\Phi _A=D_\mu ^B\Phi _B=iE_\mu I_E.
\end{array}
\end{equation}
$F$ and $E$ are a real-valued two-form and one-form on $M_3$, respectively.
Solutions to (\ref{bog'}) automatically satisfy the variational field
equations (\ref{field3}). Recalling section two, solutions to the first
equation are projectively flat connections \cite{kob}. For line bundles
this carries no extra information, but for bundles of rank greater than one
projective flatness is a strong condition. Unlike the theory of BPS magnetic
monopoles, however, the energy functional (\ref{top3}) is saturated at the
Bogomol'nyi energy with either Bogomol'nyi equation in (\ref{bog'})
satisfied. Of course the field configurations must still satisfy the
second-order variational field equations. The Bogomol'nyi energy is given by
\begin{equation}
\label{energy}{\cal E}=-\int_{M_3}(D_{[\rho }^A\Phi _A)^aH_{\mu \nu ]}^b\
{\rm tr}(T^aT^b)-\int_{M_3}(D_{[\rho }^B\Phi _B)^aK_{\mu \nu ]}^b\
{\rm tr}(T^aT^b).
\end{equation}
The added flexibility in obtaining the Bogomol'nyi energy, a feature not
present in the theory of magnetic monopoles, will lead to electric
monopoles.

Since topological monopoles, if they exist, are analogous to the BPS
magnetic monopole field configurations, we shall use the same symmetry
breaking mechanism \cite{Goddard}. We place the solitonic core region at
the origin. Let $G$ and $G_o$ be compact and connected gauge groups, where
the group $G_o$ is assumed to be embedded in $G$. The gauge group of the
core region $G$ is spontaneously broken to $G_o$ outside of the core region
when the Higgs field is covariantly constant, $D\Phi =0$. In regions far
from the core ($r\rightarrow \infty $) where we assume that
$D^A\Phi _A=0$, it can be shown that
\begin{equation}
\label{s.s.b.}H=\Phi _AF_A,
\end{equation}
where $F_A\in \Lambda ^2(M_3,E_{G_o})$, a two-form on $M_3$ taking values
in the $G_o$-Lie algebra bundle, denoted by $E_{G_o}$ here \cite{Goddard}.
An equivalent expression to (\ref{s.s.b.}) can be written when
$D^B\Phi _B=0$.
We adopt units so that $\Phi ^2=1$ when $r>>1$ and where spontaneous
symmetry breaking has occurred. When $G=U(n)$ and $G_o=U(1)$, $F_A$ becomes
a pure imaginary two-form on $M_3$. The Bogomol'nyi solitons defined by
(\ref{bog'}) have an energy coming from (\ref{energy}) topologically
fixed by
\begin{equation}
\label{bound}
\begin{array}{c}
{\cal E}=-\int_{M_3}d({\rm tr}(\Phi _AH))-\int_{M_3}d({\rm tr}(\Phi _BK))
\\ =-\int_{S^2}{\rm tr}(\Phi _AH)-\int_{S^2}{\rm tr}(\Phi _BK)
\end{array}
\end{equation}
where $S^2$ is a large sphere surrounding the monopole core. Details of
the solitonic particles in this theory now depend on the extent to which
they satisfy the Bogomol'nyi equations (\ref{bog'}) and are
`spontaneously broken'.

Consider first non-singular particle-like solutions to {\it both}
Bogomol'nyi equations in (\ref{bog'}). Substituting (\ref{s.s.b.}) into
(\ref{bound}) and using the normalization condition $\Phi ^2=1$ for both
Higgs fields, the energy is fixed by $-\int F_A-\int F_B$. Let us
conventionally interpret $\int F_A/2\pi $ as the magnetic charge ($g$),
and $\int F_B/2\pi $ as the electric charge ($q$). We can view $F_A$
and $F_B$ as curvatures on the line bundles $L_A$ and $L_B$ determined
by $\Phi _A$ and $\Phi _B$; from (\ref{s.s.b.}) $F_A$ and $F_B$ are
the projections of $H$ and $K$ on $L_A$ and $L_B$. The magnetic and
electric charges are thereby the Chern numbers associated to the
complex line bundles with curvatures $F_A$ and $F_B$, respectively, and
in this theory are thereby quantized at the classical level. The
stability of the solitonic particle is argued from the topological
interpretation, a winding number, that can be given to (\ref{bound}).
When both equations in (\ref{bog}) are satisfied the topological
stability is assured if either the electric or magnetic charge is
non-vanishing. To this point, we have used only half of the Bogomol'nyi
equations, $D^A\Phi _A=D^B\Phi _B=0$. From the projective flatness of
the curvatures in the Bogomol'nyi equations (\ref{bog'}), $H=K=FI_E$,
and (\ref{s.s.b.}) we conclude that $F=\varphi _AF_A=\varphi _BF_B$
where $\Phi_A=\varphi _AI_E$ and $\Phi _B=\varphi _BI_E$. From this
we find that
\begin{equation}
\label{dyon}\int_{S^2}F_A=\int_{S^2}F_B.
\end{equation}
Therefore non-singular, stable, particle-like solutions to the Bogomol'nyi
equations (\ref{bog'}) are dyons. To obtain electric monopoles there would
appear to be two possibilities, both resulting from a weakening of the
Bogomol'nyi equations (\ref{bog'}).

In the first possibility, one or the other of the non-abelian gauge fields,
$H$ or $K$, does not break to $U(1)$ in the far-field, so that non-abelian
gauge field passes unnoticed through the detector. This corresponds to the
case where, for example, $H^A=K^B=iFI_E=F_B\Phi _B$, but asymptotically
$D^B\Phi _B=0$, $D^A\Phi _A\neq 0$. The Bogomol'nyi bound (\ref{energy})
in this case becomes
\begin{equation}
\label{case1}{\cal E}=-\int_{S^2}F_B\ {\rm tr}(\Phi _A){\rm tr}(\Phi
_B)-\int_{S^2}F_B.
\end{equation}
The second term is $-2\pi $ times the electric charge. Since the projective
flatness reduces the gauge group to $U(n)$, we have used the Killing-Cartan
form for the bundle inner product, $<\ >$. The first integral in
(\ref{case1}) vanishes when $\Phi _A$ is traceless. An explicit example
of an electric monopole is $A=B$ projectively flat, and $\Phi _A,\Phi _B$
are any sufficiently differentiable functions on $M$ taking values in
$SU(n)$ and $U(n)$, respectively, with $\Phi _B$ covariantly constant
asymptotically ($D^B\Phi _B=0$ for $r\rightarrow \infty $). The full
variational field equations are satisfied.

The second possibility allows both gauge fields to break far from the core
using $D^A\Phi _A=D^B\Phi _B=0$, but the two gauge far-fields $H^A=F_A\Phi
_A $ and $K^B=F_B\Phi _B$ are completely decoupled so that $F_A$ and $F_B$
and their respective topological charges are unrelated. That is, the
projective flatness in the Bogomol'nyi equations is relaxed. The electric
and magnetic charges in this case become, respectively,
$$
2\pi q=-\int_{S^2}F_B,\qquad 2\pi g=-\int_{S^2}F_A,
$$
with no relation between $q$ and $g$ analogous to (\ref{dyon}).

In this section we introduced a topological field theory with a
Bogomol'nyi structure permitting BPS electric, magnetic and dyonic
monopoles. From the general arguments in \cite{olive} the particle
spectrum and classical mass compare favourably with that of the
intermediate vector bosons. Using the general relationship between
supersymmetry and the Bogomol'nyi equations it can be shown that there
are no quantum corrections to the classical mass spectrum
\cite{witoli}\cite{spec}. For $G=U(2)$, which is double covered by
$SU(2)\times U(1)$, the solitons are associated with the $W^{\pm }$
and $Z_0$ intermediate vector bosons. In this theory $W^{\pm }$ have
energies of opposite sign so that, presumably, one is the anti-particle
of the other, as required. The $Z_0$ probably corresponds to the case
where $H^A=K^B=iFI_E$, but $D^A\Phi _A\neq 0$ and $D^B\Phi _B\neq 0$.
The gauge far-fields for the $Z_0$ are thereby non-abelian and pass
unnoticed through conventional detectors, and have a stabilizing
energy given by
$$
{\cal E}=-\int_{S^2}F\ {\rm tr}(\Phi _A)-\int_{S^2}F\
{\rm tr}(\Phi _B)
$$
This completes the work begun in \cite{topdyon}.

\section{Conclusion}

We have introduced a Lagrangian field theory which contains solitonic
particles and possesses a well-behaved covariant phase space, relating
the topology of the field configuration to the dimension of the covariant
phase space. It was suggested that these classical field configurations
provide a provisional basis from which a realistic solitonic model for
force mediating bosons could be found. We have not addressed in this
paper the most important question: interaction dynamics (soliton-soliton
interactions and soliton-matter interactions).

\end{document}